\documentclass[twocolumn,aps,prl,epsf]{revtex4}
\usepackage[dvips]{graphics}
\begin{document}


\title{Anomalous Spin Response in Non-centrosymmetric Compounds}

\author{Tetsuya Takimoto}

\affiliation{Max Planck Institute for Chemical Physics of Solids, 
N$\ddot{o}$thnitzer Str. 40, 01187 Dresden, Germany}

\date{\today}

\begin{abstract}
We examine static spin susceptibilities $\chi_{\alpha\beta}({\bf q})$ 
of spin components $S_{\alpha}$ and $S_{\beta}$ 
in the non-centrosymmetric tetragonal system. 
These show anomalous momentum dependences like 
$\chi_{xx}({\bf q})-\chi_{yy}({\bf q})\sim q_x^2-q_y^2$ 
and $\chi_{xy}({\bf q})+\chi_{yx}({\bf q})\sim q_x q_y$, 
which vanish in centrosymmetric systems. 
The magnitudes of the anomalous spin susceptibilities are enhanced by 
the on-site Coulomb interaction, especially, around an ordering wave vector. 
The significant and anomalous momentum dependences of these susceptibilities 
are explained by a group theoretical analysis. 
As the direct probe of the anomalous spin susceptibility, 
we propose a polarized neutron scattering experiment. 
\end{abstract}

\maketitle

Recently, the non-centrosymmetric heavy fermion superconductor 
attracts much attention. 
For a tetragonal compound CePt$_3$Si, 
which has been intensively studied 
from both experimental and theoretical sides, 
the superconducting transition has been observed at 0.75K 
below the antiferromagnetic transition temperature 2.2K 
at ambient pressure
\cite{CePt3Si}. 
In addition, CeRhSi$_3$ and CeIrSi$_3$ also show superconductivity 
by applying pressure
\cite{CeRhSi3,CeIrSi3}. 
Commonly, superconductivity in these compounds has been 
found around the antiferromagnetic phase like in centrosymmetric 
heavy electron systems
\cite{CePt3Si,CeRhSi3,CeIrSi3,Tateiwa}. 

Theoretically, the non-centrosymmetric system is characterised by 
the Rashba-type effective spin-orbit coupling, 
which has an antisymmetric momentum dependence 
with respect to the spatial inversion 
\cite{Rashba,Frigeri1}. 
As a characteristic feature of the model, 
it has been shown that 
in terms of a band splitting caused by
the spin-orbit coupling, 
the uniform susceptibility has large Van-Vleck type contribution
\cite{Fujimoto}. 
Furthermore, the thermal average of the spin operator of an electron 
with a momentum ${\bf k}$ does not necessarily vanish
\cite{Frigeri1,Fujimoto}, 
although any magnetic moment disappears by canceling out spins 
in the paramagnetic state. 
The effect of electric spins on transport coefficients 
like spin Hall effect are intensively studied 
in the field of spintronics
\cite{Murakami,Sinova,Fujimoto}. 

For the normal state property
of non-centrosymmetric heavy fermion superconductors, 
the difference from centrosymmetric systems by the Rashba-type 
spin-orbit coupling has not been suggested except for 
the quantities mentioned above. 
Considering that in hydrostatic pressure, 
the superconductivity appears from 
the paramagnetic state with decreasing temperature, 
the spin fluctuation will be one of keys of superconductivity, and 
a normal state property 
characterising the non-centrosymmetric compound may 
relate with the mechanism of the superconductivity. 
Furthermore, as well as transport coefficients, 
it is expected that the remaining electron spins 
affect the magnetic excitation 
in the non-centrosymmetric compound. 

In this Letter, 
we examine static spin susceptibilities $\chi_{\alpha\beta}({\bf q})$ 
of $\alpha$- and $\beta$-components of spin operators, 
based on a simple Hubbard model including the Rashba-term. 
Significant momentum dependences 
like $\chi_{xx}({\bf q})-\chi_{yy}({\bf q})\sim q_x^2-q_y^2$ 
and $\chi_{xy}({\bf q})+\chi_{yx}({\bf q})\sim q_x q_y$ will be shown, 
where these susceptibilities vanish in the centrosymmetric system. 
In order to clarify the origin of momentum dependences, 
we carry out a group theoretical analysis 
to show that the symmetry of momentum dependence 
of $\chi_{\alpha\beta}({\bf q})$ is identical with 
the representation of the product of spin operators 
included in a corresponding susceptibility. 
In order to observe the unusual momentum dependence 
of spin susceptibilities, 
we suggest a polarized neutron scattering experiment, 
especially, around the magnetic instability.


In the following, we describe a non-centrosymmetric system by 
the following Hamiltonian
\begin{eqnarray}
  &&H=H_0+H_1,\\
  &&H_0=\sum_{{\bf k}\sigma\sigma'}
         [(\varepsilon_{\bf k}-\mu)\hat{\sigma}_{0}
          +{\bf g}_{\bf k}\cdot\hat{{\bf \sigma}}]_{\sigma\sigma'}
              c_{{\bf k}\sigma}^{\dagger}c_{{\bf k}\sigma'},\\
  &&H_1=U\sum_{\bf i}n_{{\bf i}\uparrow}n_{{\bf i}\downarrow},
\end{eqnarray}
where $c_{{\bf k}\sigma}$ and $c_{{\bf k}\sigma}^{\dagger}$ are 
annihilation and creation operators of an electron 
with a momentum ${\bf k}$ and a spin $\sigma$. Here, $\varepsilon_{\bf k}$ 
and $\mu$ are the energy dispersion of electrons and the chemical potential, 
respectively, 
while ${\bf g}_{\bf k}$ describes the Rashba field satisfying 
${\bf g}_{\bf k}$=$-{\bf g}_{\bf -k}$, 
which breaks the inversion symmetry. 
Then, eigenenergies of the non-interacting part are given by 
$\varepsilon_{{\bf k}\pm}=\varepsilon_{\bf k}\pm|{\bf g}_{\bf k}|-\mu$. 
In addition, $H_1$ is the on-site interaction term. 

In the non-interacting case, 
the electron Green's function is given by
\begin{eqnarray}
  \hat{G}^{(0)}({\bf k},{\rm i}\omega_n)
    =G^{(0)}_{+}({\bf k},{\rm i}\omega_n)\hat{\sigma}_{0}
    +\frac{{\bf g}_{\bf k}}{|{\bf g}_{\bf k}|}\cdot\hat{\bf \sigma}
     G^{(0)}_{-}({\bf k},{\rm i}\omega_n),
\end{eqnarray}
with
  $G^{(0)}_{\pm}({\bf k},{\rm i}\omega_n)
  =\frac{1}{2}\left[({\rm i}\omega_n-\varepsilon_{{\bf k}+})^{-1}
            \pm({\rm i}\omega_n-\varepsilon_{{\bf k}-})^{-1}\right]$,
where $\omega_n$ is a fermionic Matsubara frequency. 
Here, since $G^{(0)}_{-}({\bf k},{\rm i}\omega_n)$ 
is expanded only in odd-power of $|{\bf g}_{\bf k}|$, 
$G^{(0)}_{-}({\bf k},{\rm i}\omega_n)$ vanishes in the limit of 
$|{\bf g}_{\bf k}|\rightarrow 0$. 
Summing up the Green's function with respect to the Matsubara frequency, 
the first term gives electron densities 
of $({\bf k},\pm)$-states, 
which determine shapes of two Fermi surfaces, 
while the second term contributes electron spins 
of $({\bf k},\pm)$-states, 
whose directions are oposite from each other due to the minus sign 
in front of $({\rm i}\omega_n-\varepsilon_{{\bf k}-})^{-1}$ 
of $G^{(0)}_{-}({\bf k},{\rm i}\omega_n)$. 
These features of electrons are schematically drawn 
in Figs. 1(a) and 1(b), respectively. 
From Fig. 1(b), it is understood that spins of electrons 
belonging to $\varepsilon_{{\bf k}-}$-band remain 
in the region between two Fermi surfaces, 
while a spin of an $\varepsilon_{{\bf k}-}$-band electron 
at a ${\bf k}$ cancels out 
another spin of an $\varepsilon_{{\bf k}+}$-band electron 
at the same ${\bf k}$ 
within the inner Fermi surface. 
It is expected that the remaining electric spin affects 
the magnetic response.

\begin{figure}[t]
\resizebox{70mm}{!}
{\includegraphics{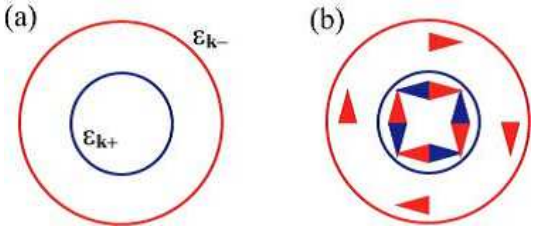}}
\caption{Schematic views of charges (a) and spins (b) of electrons. 
Red and blue lines 
denote Fermi surfaces of $\varepsilon_{{\bf k}-}$ 
and $\varepsilon_{{\bf k}+}$, respectively. Red and blue wedges 
are spins of electrons belonging to $\varepsilon_{{\bf k}-}$ 
and $\varepsilon_{{\bf k}+}$, respectively. 
Spins of $\varepsilon_{{\bf k}-}$-electrons remain 
in the region between two Fermi surfaces. }
\end{figure}


In order to examine the magnetic excitation in the non-centrosymmetic system, 
we consider the dynamical susceptibility, defined as
\begin{eqnarray}
  &&\chi_{\alpha\beta}({\bf q},{\rm i}\Omega_n)\nonumber\\
  &\hspace{-4mm}=&\hspace{-2mm}\int_0^{\frac{1}{T}}
    \hspace{-2mm}d\tau\hspace{1mm}e^{{\rm i}\Omega_n\tau}
    \langle T_{\tau}[(S^{\alpha}_{\bf q}(\tau)
                     -\langle S^{\alpha}_{\bf q}\rangle)
                     (S^{\alpha'}_{\bf -q}(0)
                     -\langle S^{\alpha'}_{\bf -q}\rangle)]\rangle, 
\label{defchi}
\end{eqnarray}
where $\langle\cdots\rangle$ means the thermal average of $\cdots$, 
$T_{\tau}$ denotes the imaginary-time chronological ordering operator, 
and $\Omega_n$ is a bosonic Matsubara frequency. 
Here, charge and spin 
operators with a wave vector ${\bf q}$ are given by 
$S^c_{\bf q}=\frac{1}{2}\sum_{{\bf k}\sigma}
c_{{\bf k}\sigma}^{\dagger}c_{{\bf k+q}\sigma}$ and 
$S^{\alpha}_{\bf q}=\frac{1}{2}\sum_{{\bf k}\sigma\sigma'}
\sigma^{\alpha}_{\sigma\sigma'}
c_{{\bf k}\sigma}^{\dagger}c_{{\bf k+q}\sigma'}$ 
($\alpha$=x, y, and z), respectively, 
where $\hat{\sigma}^{\alpha}$ is an $\alpha$-component of Pauli matrices. 
In centrosymmetric systems, 
all off-diagonal susceptibilities disappear. 
On the other hand, it should be noted that 
off-diagonal susceptibilities do not always vanish 
in non-centrosymmetric systems\cite{Yanase1}, 
and even susceptibilities between spin and charge operators 
$\chi_{\alpha c}({\bf q},{\rm i}\Omega_n)$ 
and $\chi_{c\alpha}({\bf q},{\rm i}\Omega_n)$ 
($\alpha$=x,y, and z) 
have non-zero values for $\Omega_n\neq 0$. 

According to the diagramatic technique, we formulate 
$\chi_{\alpha\beta}({\bf q},{\rm i}\Omega_n)$ with use of the Green's 
function $G^{(0)}_{\sigma\sigma'}({\bf k},{\rm i}\omega_m)$ given above. 
In the non-interacting case, the susceptibility 
$\chi^{(0)}_{\alpha\beta}({\bf q},{\rm i}\Omega_n)$ is calculated 
through a transformation from 
$\bar{\chi}_{\sigma_1\sigma_2\sigma_3\sigma_4}
({\bf q},{\rm i}\Omega_n)$ defined by, 
\begin{eqnarray}
  &&\bar{\chi}_{\sigma_1\sigma_2\sigma_3\sigma_4}
    ({\bf q},{\rm i}\Omega_n)\nonumber\\
  &\hspace{-4mm}=&\hspace{-2mm}-T\sum_{\bf k}\sum_{m}
                  G^{(0)}_{\sigma_3\sigma_1}
                   ({\bf k},{\rm i}\omega_m)
                  G^{(0)}_{\sigma_2\sigma_4}
                   ({\bf k+q},{\rm i}\omega_m+{\rm i}\Omega_n),
\end{eqnarray}
which corresponds to the one-bubble diagram
\cite{Frigeri2}. 
In the interacting case with a finite $U$, the expression of 
$\chi_{\alpha\beta}({\bf q},{\rm i}\Omega_n)$ is given within 
the random phase approximation (RPA)\cite{Yanase1,Yanase2,Tada} as
\begin{eqnarray}
 \hat{\chi}({\bf q},{\rm i}\Omega_n)=
 \left[{\bf 1}-2\hat{\chi}^{(0)}({\bf q},{\rm i}\Omega_n)\hat{U}\right]^{-1}
 \hat{\chi}^{(0)}({\bf q},{\rm i}\Omega_n),
\end{eqnarray}
where matrix elements of $\hat{U}$ are given by
$U_{\alpha\beta}=\delta_{\alpha,\beta}U_{\alpha\alpha}$ 
with $U_{cc}=-U$ and $U_{zz}=U_{xx}=U_{yy}=U$.

\begin{figure}[t]
\resizebox{75mm}{!}
{\includegraphics{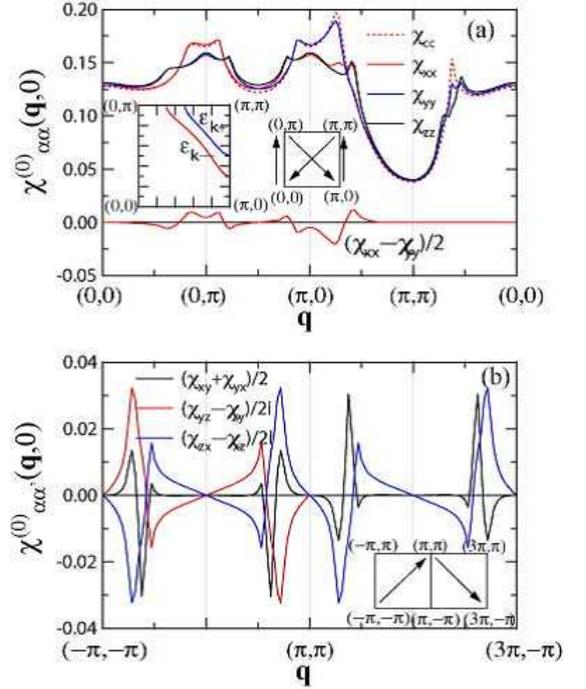}}
\caption{(a) ${\bf q}$-dependences of static susceptibilities, 
$\chi^{(0)}_{cc}({\bf q},0)$, $\chi^{(0)}_{zz}({\bf q},0)$, 
$\chi^{(0)}_{xx}({\bf q},0)$, $\chi^{(0)}_{yy}({\bf q},0)$, 
and 
$(\chi^{(0)}_{xx}({\bf q},0)-\chi^{(0)}_{yy}({\bf q},0))/2$, 
where insets show 
Fermi surfaces in the first quadrant of the Brillouin zone (left) 
and the path in ${\bf q}$-space (right). 
(b) ${\bf q}$-dependences of static susceptibilities, 
$\chi^{(0)}_{xy}({\bf q},0)$, $\chi^{(0)}_{yz}({\bf q},0)$, and 
$\chi^{(0)}_{zx}({\bf q},0)$ with the momentum path 
shown in the inset. }
\end{figure}

Then, we calculate $\chi_{\alpha\alpha'}({\bf q},{\rm i}\Omega_n)$ 
numerically, based on a two-dimensional non-centrosymmetric system 
with an energy dispersion 
$\varepsilon_{\bf k}=2t_1(\cos{k_x}+\cos{k_y})+4t_2\cos{k_x}\cos{k_y}$ 
and a Rashba-field ${\bf g}_{\bf k}=g(\sin{k_y},-\sin{k_x},0)$. 
For parameters, $t_2/t_1$=0.35 and $g/t_1$=0.2 are chosen, 
we can reproduce quasi two dimentional Fermi surfaces of CePt$_3$Si 
obtained by the band calculation\cite{Samokhin,Hashimoto,comment1}. 
In Fig. 2, we show momentum dependences of static spin susceptibilities 
in the non-interacting case. 
Diagonal components of $\hat{\chi}^{(0)}({\bf q})$ are shown in Fig. 2(a), 
where Fermi surfaces and the momentum path are also shown in insets. 
Unlike in centrosymmetric systems, all momentum dependences of 
$\chi_{cc}^{(0)}({\bf q})$, $\chi_{zz}^{(0)}({\bf q})$, 
$\chi_{xx}^{(0)}({\bf q})$, and $\chi_{yy}^{(0)}({\bf q})$ are different 
from each other. 
Especially, in a path from (0,$\pi$) to ($\pi$,0), 
$\chi_{xx}^{(0)}({\bf q})-\chi_{yy}^{(0)}({\bf q})$ is antisymmetric 
around the mid-point ($\pi$/2,$\pi$/2), and 
it vanishes in a diagonal path from ($\pi$,$\pi$) to (0,0). 
Therefore, the momentum dependence of 
$\chi_{xx}^{(0)}({\bf q})-\chi_{yy}^{(0)}({\bf q})$ 
has the typical $q_x^2-q_y^2$ symmetry. 
Likewise, momentum dependences of 
$\chi_{xy}^{(0)}({\bf q})+\chi_{yx}^{(0)}({\bf q})$, 
$(\chi_{yz}^{(0)}({\bf q})-\chi_{zy}^{(0)}({\bf q}))/{\rm i}$, 
and $(\chi_{zx}^{(0)}({\bf q})-\chi_{xz}^{(0)}({\bf q}))/{\rm i}$ are 
shown in Fig. 2(b), where the momentum path is depicted in the inset. 
In the figure, 
$\chi_{xy}^{(0)}({\bf q})+\chi_{yx}^{(0)}({\bf q})$ is symmetric 
around (0,0) and (2$\pi$,0), 
while it is antisymmetric around ($\pi$,$\pi$). 
Accordingly, the momentum dependence of 
$\chi_{xy}^{(0)}({\bf q})+\chi_{yx}^{(0)}({\bf q})$ 
is a $q_x q_y$-type. 
Similarly, it can be understood easily that 
momentum dependences of 
$(\chi_{yz}^{(0)}({\bf q})-\chi_{zy}^{(0)}({\bf q}))/{\rm i}$ 
and $(\chi_{zx}^{(0)}({\bf q})-\chi_{xz}^{(0)}({\bf q}))/{\rm i}$ are 
$q_y$- and $q_x$-types, respectively. 
Thus, spin susceptibilities 
have significant and unusual momentum dependeces 
related with corresponding spin indices.

\begin{figure}[t]
\resizebox{75mm}{!}
{\includegraphics{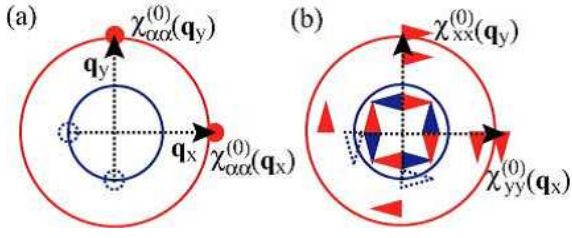}}
\caption{Schematic views of particle-hole excitations. 
Red and blue lines denote Fermi surfaces of $\varepsilon_{{\bf k}-}$ 
and $\varepsilon_{{\bf k}+}$, respectively. Red and blue wedges 
are spins of electrons belonging to $\varepsilon_{{\bf k}-}$ 
and $\varepsilon_{{\bf k}+}$, respectively. 
In (b), particle-hole excitations with momenta 
${\bf q}_x$ and ${\bf q}_y$ 
only contribute to $\chi_{yy}^{(0)}({\bf q}_x)$ and 
$\chi_{xx}^{(0)}({\bf q}_y)$, respectively, 
while every particle-hole excitaion is isotropic 
for the spin index $\alpha$ in (a).}
\end{figure}

For an intuitive picture of anomalous momentum dependences of 
spin susceptibilities, 
we consider particle-hole excitations in the non-centrosymmetric system. 
According to the prescription to calculate 
$\chi_{\alpha\beta}^{(0)}({\bf q})$, 
contributions to $\chi_{xx}^{(0)}({\bf q})$ and $\chi_{yy}^{(0)}({\bf q})$ 
are separated to two terms of convolutions, 
$T\sum_{m}G^{(0)}_{+}({\bf k},{\rm i}\omega_m)
G^{(0)}_{+}({\bf k+q},{\rm i}\omega_m+{\rm i}\Omega_n)$ and 
$T\sum_{m}G^{(0)}_{-}({\bf k},{\rm i}\omega_m)
G^{(0)}_{-}({\bf k+q},{\rm i}\omega_m+{\rm i}\Omega_n)$, 
where the second part vanishes in the limit of the zero Rashba field. 
Here, we show schematic figures of these particle-hole excitations 
in Figs. 3(a) and 3(b), respectively. 
In Fig. 3(a), the mechanism of particle-hole excitations 
is not different from that of the centrosymmetric case. 
In this case, 
the contribution to $\chi_{\alpha\alpha}^{(0)}({\bf q})$ 
is isotropic in the spin space. 
On the other hand, a particle-hole excitation 
with a momentum transfer ${\bf q}_x$ shown in Fig. 3(b) 
only contributes to $\chi_{yy}^{(0)}({\bf q}_x)$ for low energy excitations: 
rotating the quantization axis to the $-{\bf e}_y$ direction, 
the longitudinal spin excitation is only permitted, 
and other excitations are almost forbidden 
because of the energy loss from spin rotations against the Rashba-field. 
Similarly, if we consider a particle-hole excitation 
with a momentum transfer ${\bf q}_y$ shown in Fig. 3(b), 
only $\chi_{xx}^{(0)}({\bf q}_y)$ has the contribution. 
This is consistent with a behavior in a path (0,0)-(0,$\pi$) of Fig. 2(a), 
where the momentum dependence of $\chi_{xx}^{(0)}({\bf q})$ is 
considerably different from those of 
$\chi_{zz}^{(0)}({\bf q})$ and $\chi_{yy}^{(0)}({\bf q})$. 
Considering these points, 
$\chi_{xx}^{(0)}({\bf q})-\chi_{yy}^{(0)}({\bf q})$ should have 
the $q_x^2-q_y^2$-type momentum dependence. 
Similar consideration will also work well for 
$\chi_{xy}^{(0)}({\bf q})+\chi_{yx}^{(0)}({\bf q})$.

\begin{figure}[t]
\resizebox{70mm}{!}
{\includegraphics{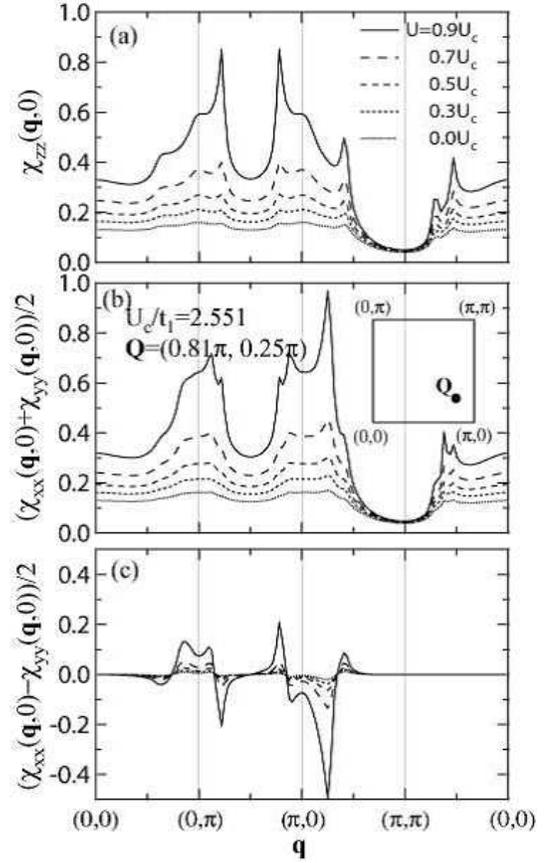}}
\caption{${\bf q}$-dependences of static susceptibilities 
(a) $\chi_{zz}({\bf q},0)$, 
(b) $(\chi_{xx}({\bf q},0)+\chi_{yy}({\bf q},0))/2$, and 
(c) $(\chi_{xx}({\bf q},0)-\chi_{yy}({\bf q},0))/2$
for several values of $U$. The position of the ordering wave vector 
is depicted in the inset of (b). }
\end{figure}

The above discussion is for the non-interacting case. 
By the interaction $U$, magnitudes of 
anomaous $\chi_{xx}({\bf q})-\chi_{yy}({\bf q})$ and 
$\chi_{xy}({\bf q})+\chi_{yx}({\bf q})$ 
should be changed. 
In order to investigate the effect of interaction on 
${\bf q}$-dependences of susceptibilities, 
we calculate $\chi_{\alpha\beta}({\bf q})$ within RPA 
using the same parameter set as in Fig. 2. 
With the critical interaction $U_{\rm c}=2.551t_1$, 
the paramagnetic state becomes unstable 
at an ordering wave vector ${\bf Q}=(0.81\pi,0.25\pi)$. 
We show momentum dependences of $\chi_{zz}({\bf q})$, 
$(\chi_{xx}({\bf q})+\chi_{yy}({\bf q}))/2$, and 
anomalous $(\chi_{xx}({\bf q})-\chi_{yy}({\bf q}))/2$ 
in Figs. 4(a), 4(b), and 4(c), respectively. 
$\chi_{zz}({\bf q})$ and
$(\chi_{xx}({\bf q})+\chi_{yy}({\bf q}))/2$ are enhanced 
with increasing $U$, but have different ${\bf q}$-dependences 
from each other. 
Furthermore, the amplitude of 
$(\chi_{xx}({\bf q})-\chi_{yy}({\bf q}))/2$ increases considerably 
around the ordering wave vector ${\bf Q}$ in comparison with 
the corresponding value of the non-interacting case. 
In addition, we note that the amplitude of 
$\chi_{xy}({\bf q})+\chi_{yx}({\bf q})$ is also 
enhanced with increasing $U$.

\begin{table}[t]
\caption{Classification of spin product $S^{\alpha}S^{\beta}$ 
according to irreducible representations 
of the C$_{\rm 4v}\times{\mathcal K}$ group. 
Then, $S^{\rm c}$, $S^{\rm z}$, and $\{S^{\rm x},S^{\rm y}\}$ belong to 
A$_1^+$, A$_2^-$, and E$^-$ irreducible representations, respectively. 
The third column shows the basis function in the momentum space. 
The superscript + (-) of irreducible representations 
expresses the even (odd) parity with respect to the time reversal. }
\begin{center}
\begin{tabular}{ccc}
$\Gamma$ & $S^{\alpha}S^{\beta}$ & basis\\
\hline
A$_1^+$ & $S^{\rm c}S^{\rm c}$ & $q_z^2$, $q_x^2+q_y^2$\\
 & $S^{\rm z}S^{\rm z}$ &\\
 & $S^{\rm x}S^{\rm x}+S^{\rm y}S^{\rm y}$ &\\
A$_2^+$ & i$(S^{\rm c}S^{\rm z}-S^{\rm z}S^{\rm c})$ 
              & $q_xq_y(q_x^2-q_y^2)$\\
B$_1^+$ & $S^{\rm x}S^{\rm x}-S^{\rm y}S^{\rm y}$ 
              & $q_x^2-q_y^2$\\
B$_2^+$ & $S^{\rm x}S^{\rm y}+S^{\rm y}S^{\rm x}$ 
              & $q_xq_y$\\
E$^+$ & $\{S^{\rm y}S^{\rm z}+S^{\rm z}S^{\rm y}$, 
              $S^{\rm z}S^{\rm x}+S^{\rm x}S^{\rm z}\}$ 
        & $\{q_yq_z, q_zq_x\}$\\
      & $\{$i$(S^{\rm c}S^{\rm x}-S^{\rm x}S^{\rm c})$, 
              i$(S^{\rm c}S^{\rm y}-S^{\rm y}S^{\rm c})\}$ &\\
A$_2^-$ & $S^{\rm c}S^{\rm z}+S^{\rm z}S^{\rm c}$ 
          & $q_xq_yq_z(q_x^2-q_y^2)$\\
        & i$(S^{\rm x}S^{\rm y}-S^{\rm y}S^{\rm x})$ &\\
E$^-$ & $\{$i$(S^{\rm y}S^{\rm z}-S^{\rm z}S^{\rm y})$, 
              i$(S^{\rm z}S^{\rm x}-S^{\rm x}S^{\rm z})\}$ 
        & $\{q_x, q_y\}$\\
      & $\{S^{\rm c}S^{\rm x}+S^{\rm x}S^{\rm c}$, 
              $S^{\rm c}S^{\rm y}+S^{\rm y}S^{\rm c}\}$ &\\
\hline
\end{tabular}
\end{center}
\label{d4h}
\end{table}

In order to clarify the origin of the unusual momentum dependence 
of the spin susceptibility, we carry out a group theoretical analysis. 
In the present case, the relevant group is C$_{\rm 4v}\times{\mathcal K}$, 
where C$_{\rm 4v}$ is the tetragonal point group without the inversion 
symmetry, and ${\mathcal K}$ is the time-reversal symmetry group. 
From the definition, 
the momentum dependence of 
$\chi_{\alpha\beta}({\bf q},{\rm i}\Omega_n)$ is determined by 
the expectation value of the right hand side. 
Noting that both spin and momentum are transformed 
by symmetric operations of the group, 
the representation of the operator in the expectation value 
is $\Gamma_{s}\otimes\Gamma_{m}$, 
where $\Gamma_s$ and $\Gamma_m$ are representations of the spin product 
and the momentum dependence, respectively. 
Since the operator giving a non-zero 
expectation value $\langle\hat{O}\rangle$ belongs to 
the identity representation $\Gamma_1$ (A$_1^+$ in Table I) in the group, 
the relation $\Gamma_{s}\otimes\Gamma_{m}=\Gamma_1$ is obtained, 
so that finally 
\begin{eqnarray}
  \Gamma_{m}=\Gamma_{s}
\end{eqnarray}
is required. 
For the symmetry of the ${\bf q}$-dependence 
of $\chi_{\alpha\beta}({\bf q},{\rm i}\Omega_n)$, 
it is sufficient to know the representation of the spin-product 
$S^{\alpha}S^{\beta}$, 
where the classification of the product is given in Table I. 
Therefore, 
representations of ${\bf q}$-dependences of
$\chi_{xx}({\bf q})-\chi_{yy}({\bf q})$ 
and $\chi_{xy}({\bf q})+\chi_{yx}({\bf q})$ are B$_1^+$ and B$_2^+$, 
respectively. 
These results are consistent with ${\bf q}$-dependences of 
anomalous susceptibilities shown in Figs. 2 and 4. 
Thus, it is shown that 
the symmetry of the ${\bf q}$-dependence 
of $\chi_{\alpha\beta}({\bf q},{\rm i}\Omega_n)$ 
is identical with the representation of the spin-product 
$S^{\alpha}S^{\beta}$. 

Finally, we suggest a polarized neutron scattering experiment 
to observe the novel anomalous momentum dependence of susceptibility 
in non-centrosymmetric heavy fermion systems, 
where even anomalous susceptibilities will be enhanced 
by the strong interaction as shown in Fig. 4. 
Among non-centrosymmetric heavy fermion compounds, 
the only available neutron scattering data are 
for magnetic structures of CePt$_3$Si\cite{Metoki} 
and CeRhSi$_3$\cite{Aso}. 
For CePt$_3$Si, considering the ordering wave vector (0,0,$\pi$), 
to observe the anomalous momentum dependence of susceptibilities 
will be difficult, 
because the symmetry of ${\bf q}$-dependence of 
every anomalous susceptibility is not the A$_1^+$ representation. 
On the other hand, for CeRhSi$_3$, 
whose ordering wave vector is ($\pm$0.43$\pi$,0,$\pi$), 
it will be more promising to observe the anomalous momentum dependence 
of, especially, $\chi_{xx}({\bf q})-\chi_{yy}({\bf q})$ 
just above the magnetic transition temperature, 
since the susceptibility 
will be enhanced around the ordering wave vector 
by approaching the magnetic transition temperature. 

In summary, we have examined the dynamical susceptibility 
in the non-centrosymmetric tetragonal system. 
It has been shown that affected by the Rashba term, 
$\chi_{xx}({\bf q})-\chi_{yy}({\bf q})$ and 
$\chi_{xy}({\bf q})+\chi_{yx}({\bf q})$ 
have $q_x^2-q_y^2$- and $q_x q_y$-types of momentum dependences, 
respectively. 
The group theoretical analysis has been used to explain 
the peculiar feature in the non-centrosymmetric systems. 
Since the unusual susceptibility is also enhanced by the Coulomb interaction, 
it is desirable to observe the unusual momentum dependence of susceptibility 
by the polarized neutron scattering experiment. 

The author is grateful to P. Thalmeier 
for many valuable discussions. 

\vskip-0.5cm


\begin{thebibliography}{99}
\bibitem{comment1}
If we will discuss CePt$_3$Si realistically, 
we should use three dimensional energy dispersion 
to obtain appropriate ordering wave vector (0,0,$\pi$).
\end{thebibliography}
\end{document}